\documentclass[english,preprintnumbers]{revtex4}

\usepackage{graphicx}
\usepackage{babel}
\usepackage{citesort}

\begin{document}

\preprint{P5\_nadolsky\_0710}

\preprint{ANL-HEP-CP-01-105}

\preprint{FERMILAB-Conf-01/330-T}

\preprint{MSUHEP-11025}

\preprint{SMU-HEP 01/11}

\title{PDF uncertainties in \( WH \) production at Tevatron}

\author{P.M. Nadolsky\protect\( ^{1,2} \)}

\email{nadolsky@pa.msu.edu}

\thanks{work supported by the National Science Foundation, U. S. Department of
        Energy, and Lightner-Sams Foundation}

\author{Z. Sullivan\protect\( ^{3,4} \)}

\email{zack@fnal.gov}

\thanks{work at Argonne supported by the U. S. Department of Energy
under contract No. W-31-109-ENG-38}

\thanks{work at Fermilab supported by the U. S. Department of Energy
under contract No. DE-AC02-76CH03000.}

\affiliation{\( ^{1} \)Department of Physics \& Astronomy,\nolinebreak[4] Michigan State
University, East Lansing, MI, 48824\\
 \( ^{2} \)Department of Physics, Southern Methodist University, Dallas,
TX, 75275\\
 \( ^{3} \)High Energy Physics Division, Argonne National Laboratory, Argonne,
IL, 60439\\
 \( ^{4} \)Theoretical Physics Department, Fermi National Accelerator Laboratory,
Batavia, IL, 60510-0500}

\date{29 October, 2001}

\begin{abstract}
We apply a method proposed by members of CTEQ Collaboration to
estimate the uncertainty in associated $W$-Higgs boson production at Run
II of the Tevatron due to our imprecise knowledge of parton
distribution functions. We find that the PDF uncertainties for the
signal and background rates are of the order 3\%. The PDF
uncertainties for the important statistical quantities (significance
of the Higgs boson discovery, accuracy of the measurement of the \( WH
\) cross section) are smaller (1.5\%) due to the strong correlation of
the signal and background.
\end{abstract}
\maketitle

The steady improvement of world hadronic data has stimulated
significant interest in quantitative estimates of theoretical
uncertainties due to incomplete knowledge of parton distribution
functions (PDFs) [1-7].  The
published studies discuss the impact of uncertainties on the shape of
the PDFs or simple observables, such as the total cross section for
the \( W- \)boson production at Tevatron. It is interesting to apply
the proposed methods to more involved observables, such as cross
section distributions or the ratio of the signal to background in the
search for new physics.

To illustrate new issues that such application involves, consider how
PDF uncertainties affect the potential of discovery of the Standard
Model Higgs boson with the mass \( M_{H}=115 \) GeV via associated \(
WH \) production at Tevatron. In this process, Higgs bosons can be
discovered by observing excess production of \( b\bar{b} \) pairs with
the invariant mass close to \( M_{H} \), \emph{e.g.,} in the band \(
95\leq M_{b\bar{b}}\leq 135 \) GeV. The \( WH \) signal consists of
two tagged \( b \)-quark jets, a lepton, and missing transverse energy
associated with the unobserved neutrino from the \( W \) decay. The
dominant background process is direct \( Wb\bar{b} \) production. In
an experimental analysis, this QCD background would be estimated by
extrapolation of the \( M_{b\bar{b}} \) distribution from the regions
of \( M_{b\bar{b}} \) where the \( WH \) cross section is negligibly
small, \emph{e.g.}, from the side bands \( 75\leq M_{b\bar{b}}\leq
95\mbox {\, GeV} \) and \( 135\leq M_{b\bar{b}}\leq 155 \) GeV.

We model both the signal and background with tree-level matrix element
calculations using MADGRAPH~\cite{Stelzer:1994ta} at a hard scattering
scale \( \mu ^{2}=\hat{s} \).  To simulate the resolution of the
hadron calorimeter, we smear the jet energies with a Gaussian of width
\( \Delta E_{j}/E_{j}=0.80/\sqrt{E_{j}}\oplus 0.05 \) (added in
quadrature). We simulate the acceptance of the detector by using the
selections listed in Table~\ref{P5_nadolsky_0710acc}. An isolation cut is
placed on the lepton, as defined by a cone of radius \( \Delta R
\). We use the impact-parameter \( b \)-tagging efficiency function
defined in SHW \( 2.3 \) \cite{Carena:2000yx}.
\begin{table}
\vspace*{-2em}
\caption{Cuts used to simulate the acceptance of the detector at the
Tevatron run~II.
\label{P5_nadolsky_0710acc}}
{\centering \begin{tabular}{ll}
 \( |\eta _{b}|<2 \)&
 \( E_{Tb}>20 \) GeV \\
 \( |\eta _{l}|<1.5 \)&
 \( E_{Tl}>20 \) GeV \\
 \( |\Delta R_{bl}|>0.7 \)&
 \( {\not \! E}_{T}>20 \) GeV \\
\end{tabular}}
\vspace*{-2em}
\end{table}

The PDF uncertainties influence the potential for the discovery of the
Higgs boson in several ways. First, they affect the shape of the \(
M_{b\bar{b}} \) distribution and, therefore, the accuracy of the
extrapolation of the background from the side bands. Second, the
relative errors for statistical quantities, such as the ratio \( S/B
\) of the signal and background rates, may differ significantly from
the relative errors for \( S \) and \( B \) if the latter ones are
correlated or anticorrelated. Third, the errors for \( S \)
and \( B \) are likely to be different in the positive and negative
directions, commonly due to the changes in the shape of the
distributions \( d\sigma /dM_{b\bar{b}} \) under the variation of the
PDFs. Hence, the integrated distribution may be less constrained in
the positive direction than in the negative direction.

As a first step in the study of the above issues, we estimate the PDF
uncertainties for \( S \) and \( B \) with the method proposed by
J. Pumplin, D. Stump, Wu-Ki Tung \emph{et al.} (PST)
\cite{Pumplin:2001ct}. The PST method is based on diagonalization of
the matrix of second derivatives for \( \chi ^{2} \) (Hessian matrix)
near the minimum of \( \chi ^{2} \). Since \( \chi ^{2} \) is
approximately parabolic near its minimum \( \chi ^{2}_{0} \),
hypersurfaces of constant \( \chi ^{2} \) are hyperellipses in the
space of the original 16 PDF parameters \( \{a_{i}\} \). By an
appropriate change of coordinates \( \{a_{i}\}\rightarrow \{z_{i}\},\,
i=1,\dots ,16 \), we can transform hyperellipses into hyperspheres. We
assume that all acceptable PDF sets correspond to \( \chi ^{2} \) that
does not exceed its minimal value \( \chi _{0}^{2} \) more than by \(
T^{2} \). As a result, the acceptable PDF sets have \( \{z_{i}\} \)
within a sphere of the radius \( T^{2} \) around \( \{z_{i}(\chi
_{0}^{2})\}\equiv \{z_{i}^{0}\} \).  We present the results for \(
T=10 \). Our global analysis of the PDFs uses the same set of hadronic
data as in Ref.~\cite{Pumplin:2001ct}.

\setlength{\textheight}{241mm}
The PDF uncertainty for an observable \( O \) is the maximal change in
\( O \) as a function of variables \( \{z_{i}\} \) varying within the
tolerance hypersphere. The PST method estimates the variation of \( O
\) as
\begin{equation}
\label{P5_nadolsky_0710dO}
\delta O=\sqrt{\sum ^{16}_{i=1}\delta O_{i}^{2}},\mbox {\, where\, }\delta O_{i}\equiv T\frac{\partial O}{\partial z_{i}}\approx T\frac{O(z^{0}_{i}+t)-O(z^{0}_{i}-t)}{2t},
\end{equation}
 and \( t=5 \) is a small step in the space of \( z_{i} \). For brevity
\( O(z_{1}^{0},\dots ,z_{i}^{0}\pm t,\dots ,z_{16}^{0}) \) is denoted as
\( O(z_{i}^{0}\pm t) \).

The PDF error (\ref{P5_nadolsky_0710dO}) is a combination of 32 cross sections, each of
which is known with some uncertainty due to the Monte-Carlo integration.
One might be concerned about accumulation of Monte-Carlo errors in the process
of calculation of \( \delta O \). Fortunately, this accumulation does not
happen, because the calculation of \( \delta O \) in Eq.~(\ref{P5_nadolsky_0710dO}) and
the propagation of Monte-Carlo errors involves only summation in quadrature.
The Monte-Carlo error \( \Delta _{MC}\delta O \) for \( \delta O \) is
given by\begin{equation}
\label{P5_nadolsky_0710ddS2}
\Delta _{MC}\delta O=\frac{1}{\delta O}\left( \frac{T}{2t}\right) ^{2}\sqrt{\sum _{i=1}^{16}\left( O_{i}-O_{i+1}\right) ^{2}(\Delta _{MC}O_{i}^{2}+\Delta _{MC}O_{i+1}^{2})},
\end{equation}
 where \( \Delta _{MC}O_{i} \) are Monte-Carlo errors for \( O \) calculated
with the PDF set \( i \). If all \( \Delta _{MC}O_{i} \) are approximately
the same (\( \Delta _{MC}O_{i}\approx \Delta  \) for \( i=1,\dots \, 32 \)),
Eq.~(\ref{P5_nadolsky_0710ddS2}) simplifies to\begin{equation}
\label{P5_nadolsky_0710ddO}
\Delta _{MC}\delta O\approx \frac{T}{2t}\Delta \sqrt{2}.
\end{equation}
 Hence, in this case the Monte-Carlo error for \( \delta O \) is proportional
to the Monte-Carlo error for \( O_{i} \) and not to the number of the PDF
parameters. In our calculation, the Monte-Carlo uncertainty of \( \delta O \)
does not exceed 20\% of \( \delta O. \)

In addition to the symmetric error \( \delta O \), it is useful to estimate
maximal variations of \( O \) in the positive and negative directions, given
by\begin{eqnarray}
\delta O_{-} & = & \frac{T}{t}\sqrt{\sum _{i=1}^{16}\left[ \max \left( O(z_{i}^{0})-O(z^{0}_{i}+t),O(z_{i}^{0})-O(z^{0}_{i}-t),0\right) \right] ^{2}},\label{P5_nadolsky_0710dOminus} \\
\delta O_{+} & = & \frac{T}{t}\sqrt{\sum _{i=1}^{16}\left[ \max \left( O(z^{0}_{i}+t)-O(z_{i}^{0}),O(z^{0}_{i}-t)-O(z_{i}^{0}),0\right) \right] ^{2}}.\label{P5_nadolsky_0710dOplus} 
\end{eqnarray}
 These variations define the true allowed range for \( O \) and may differ
significantly from \( \delta O \). For instance, \( \delta O \) for the
total cross section of \( W^{\pm } \)-boson at the LHC is \( 5\% \), while
\( \delta O_{-} \) and \( \delta O_{+} \) are \( -3 \) and \( +7\% \),
respectively.

In the associated \( WH \) production at the integrated luminosity \( \int Ldt=15\mbox {\, fb}^{-1} \),
we expect the following numbers of the signal and background events in the
signal band \( 95\leq M_{b\bar{b}}\leq 135\mbox {\, GeV} \):\begin{eqnarray}
S=49.7_{-1.4}^{+1.8}\, (3.0\%), &  & B=110.1_{-4}^{+3.6}\, (3.1\%).\label{P5_nadolsky_0710SB1} 
\end{eqnarray}
 The number in parentheses is the symmetric relative error for \( S \) or
\( B \) estimated with the help of Eq.~(\ref{P5_nadolsky_0710dO}). Common statistical
combinations of \( S \) and \( B \) are\begin{eqnarray}
S/B=0.451_{-0.006}^{+0.011}\, (1.5\%), &  & S/\sqrt{B}=4.73_{-0.07}^{+0.13}(1.8\%),\label{P5_nadolsky_0710SB2} \\
\sqrt{S+B}/S & = & 0.254_{-0.006}^{+0.004}(1.7\%).\label{P5_nadolsky_0710SB3} 
\end{eqnarray}
 According to Eqs.~(\ref{P5_nadolsky_0710SB2}-\ref{P5_nadolsky_0710SB3}), the errors for the statistical
quantities are very asymmetric. In the lower (\( 75\leq M_{b\bar{b}}\leq 95 \)
GeV) and upper (\( 135\leq M_{b\bar{b}}\leq 155\mbox {\, GeV} \)) side bands,
we expect \( 94.5_{-5.7}^{+2.3}(4\%) \) and \( 30_{-0.06}^{+3.3}(5\%) \)
background events, respectively.

\begin{figure}[tbh]
\mbox{
\begin{minipage}{0.48\textwidth}\vspace*{-2em}
\centerline{\vspace*{-1em}\resizebox*{7.3cm}{!}{\includegraphics{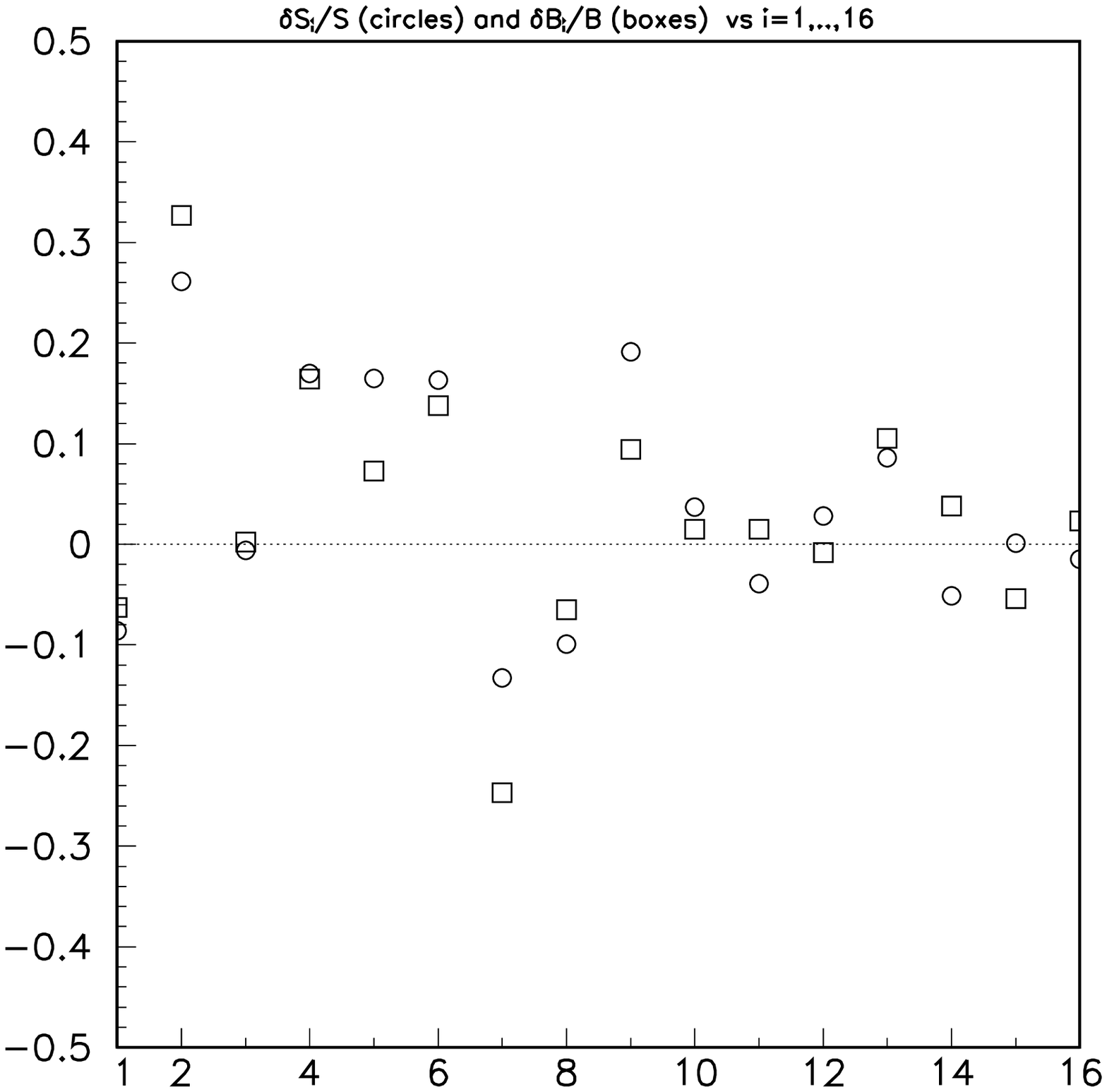}}}
\caption{Spread of $\delta S_{i}/S$ (circles) and $\delta B_{i}/B$
(boxes) for $i=1,\dots ,16$.}
\label{P5_nadolsky_0710f1}
\end{minipage}\hspace*{0.02\textwidth}
\begin{minipage}{0.48\textwidth}
\centerline{\vspace*{-1em}\resizebox*{7.3cm}{!}{\includegraphics{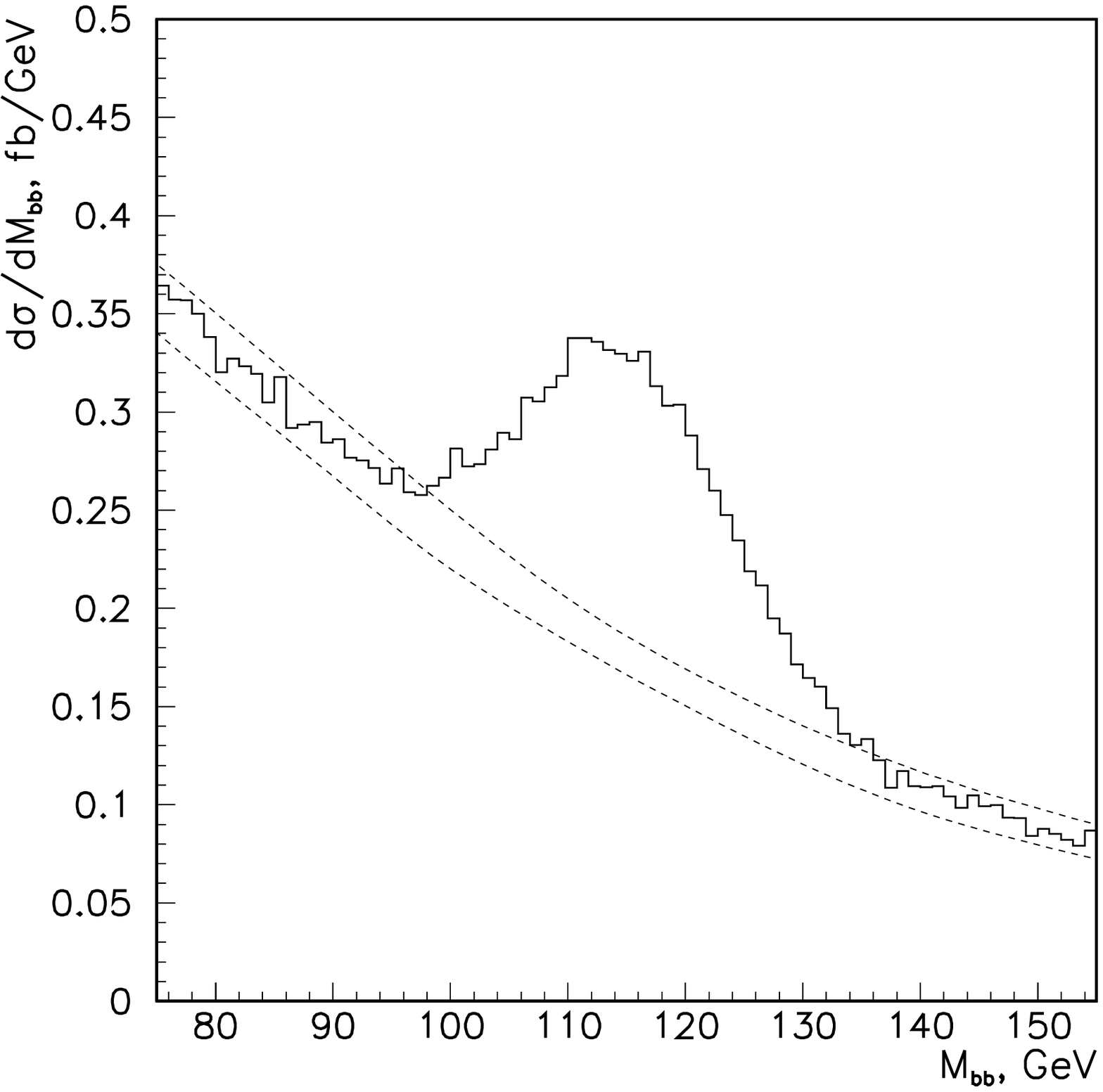}}}
\caption{The total signal plus background $M_{b\bar{b}}$ distribution (solid
line) as compared to the PDF uncertainty band for the background (dashed
lines).}
\label{P5_nadolsky_0710f2}
\end{minipage}
}\vspace*{-1.5em}
\end{figure}
%{\centering \vspace*{-1em}\resizebox*{7.3cm}{!}{\includegraphics{ds_s_db_b.eps}} \resizebox*{7.3cm}{!}{\includegraphics{sb.eps}} \par}

Eqs.~(\ref{P5_nadolsky_0710SB1}-\ref{P5_nadolsky_0710SB3}) also show that the PDF errors on \( S/B,\,
S/\sqrt{B},\sqrt{S+B}/S \) are smaller (1.5\%-1.8\%) than the
uncertainties on \( S \) and \( B \) (\( \sim 3\% \)), which signals a
correlation between the PDF errors for \( S \) and \( B \). We can get
a feeling of this correlation by studying correlations of individual
variations \( \delta S_{i}/S \) and \( \delta B_{i}/B \)
(Fig.~\ref{P5_nadolsky_0710f1}). On average, the magnitude of \(
\delta S_{i}/S \) is larger than the magnitude of corresponding \(
\delta B_{i}/B \), since the {}``average ratio{}'' of these magnitudes
\( \frac{1}{16}\sum _{i=1}^{16}|(\delta S_{i}/S)/(\delta
B_{i}/B)|=1.53 \) exceeds unity. However, we are more interested in
the correlation of the largest values of \( \delta S_{i}/S \) and \(
\delta B_{i}/B \), which give dominant contributions to the total
relative errors of \( S,\, B, \) and statistical quantities. According
to Fig.~\ref{P5_nadolsky_0710f1}, the largest \( \delta S_{i}/S \) and
\( \delta B_{i}/B \) are well correlated, so that their contributions
to \( \delta (S/B)/(S/B) \), etc., cancel. As a result, the
relative errors for the statistical quantities are smaller than the
relative errors for \( S \) and \( B. \)

The overall correlation of vectors \( \{\delta S_{i}/S\} \) and \(
\{\delta B_{i}/B\} \) can be quantified by introducing the cosine of
the angle between these vectors \cite{Pumplin:2001ct}:
\begin{equation}
\cos \varphi \equiv \frac{1}{\delta S\, \delta B}\sum _{i=1}^{16}\delta S_{i}\delta B_{i}.
\end{equation}
Then, the relative error for \( A\equiv S/B^{p} \) is
\begin{equation}
\left( \frac{\delta A}{A}\right) ^{2}=\left( \frac{\delta S}{S}\right) ^{2}+\left( \frac{\delta B^{p}}{B^{p}}\right) ^{2}-2\frac{\delta S}{S}\frac{\delta B^{p}}{B^{p}}\cos \varphi ,
\end{equation}
and correlated or anticorrelated \( S \) and \( B \) correspond to \(
\cos \varphi =1 \) or \( -1 \), respectively. Similar correlation
angles can be calculated for any pair of relative errors, including
the errors for backgrounds in the upper and lower side bands. We find
that \( \cos \varphi \) for \( S \) and \( B \), \( S+B \) and \( S \)
in the signal band are \( 0.89 \), \( 0.95 \), respectively,
\emph{i.e.}, the correlation is very good. The correlation cosine
between the background cross sections in the lower and upper side
bands is also large (\( 0.62 \)), which indicates that the PDF
uncertainty mostly affects the overall normalization of the \(
M_{b\bar{b}} \) distribution and not its shape. Correspondingly, the
extrapolation from the side bands accurately approximates the
background in the signal band. Fig.~\ref{P5_nadolsky_0710f2} illustrates the
relative size of PDF uncertainties for the background in comparison to
the signal plus background distribution.

To conclude, we propose to use asymmetric PDF errors and correlations
between PDF errors in detailed studies of PDF uncertainties. Using
these quantities, we find that the cumulative effect of the PDF
uncertainties on the significance for the discovery of the Higgs
bosons at Tevatron is not large (\( \sim 1.5-1.8\% \)).  The PDF
uncertainties for the signal and background (\( \sim 3\% \)) are much
smaller than the eventual statistical errors for the measurement of
the \( WH \) cross section (\( \gtrsim 25\% \)) even if Tevatron Run
II accumulates \( 15\mbox {\, fb}^{-1} \) of the integrated
luminosity.

\end{document}